\documentclass[pre,twocolumn,showpacs,showkeys,preprintnumbers,floatfix]{revtex4-1}

\usepackage{etex}
\usepackage{ifpdf}
\usepackage{hyperref}
\usepackage{dcolumn}
\usepackage{url}
\usepackage{amsmath}
\usepackage{amscd}
\usepackage{amsfonts}
\usepackage{amssymb}
\usepackage{amsthm}
\usepackage{bigints}
\usepackage{xfrac}
\usepackage{braket}
\usepackage{bm}   
\usepackage{bbm}
\usepackage{verbatim}
\usepackage{stmaryrd}
\usepackage{xcolor}
\usepackage{setspace}
\usepackage{tikz}
\usetikzlibrary{arrows}
\usetikzlibrary{automata}
\usetikzlibrary{positioning}
\usetikzlibrary{plotmarks}
\usetikzlibrary{calc}
\usepackage{pgfplots}
\pgfplotsset{compat=newest}

\usepackage{hyperref}
\hypersetup{
    colorlinks,
    citecolor=blue,
    filecolor=blue,
    linkcolor=blue,
    urlcolor=blue
}

\definecolor{commentred}{rgb}{0.8, 0.2, 0.2}
\definecolor{commentgreen}{rgb}{0.2, 0.8, 0.2}

\newcommand{\Abet}{\ProcessAlphabet}
\newcommand{\MS}{\MeasSymbol}
\newcommand{\MSs}{\MeasSymbols}
\newcommand{\ms}{\meassymbol}

\newcommand{\St}{\CausalState}
\newcommand{\st}{\causalstate}

\renewcommand{\H}{\operatorname{H}}
\newcommand{\I}{\operatorname{I}}

\tikzstyle{vaucanson}=[
  node distance=2.5cm,
  bend angle=15,
  auto,
  every loop/.style={},
  every edge/.style={->,draw=black,line width=1.2,>=latex,shorten <=1pt,shorten >=1pt},
  every state/.style={draw=black,line width=2,font=\large},
  loop right/.style={right,out=22,in=-22,loop},
  loop above/.style={above,out=112,in=68,loop},
  loop left/.style={left,out=202,in=158,loop},
  loop below/.style={below,out=292,in=248,loop},
  loop above right/.style={right,out=67,in=23,loop},
  loop above left/.style={left,out=157,in=113,loop},
  loop below left/.style={left,out=247,in=203,loop},
  loop below right/.style={right,out=337,in=293,loop},
]

\tikzset{Pstate/.style={state, draw=red},
	NPstate/.style={state, draw=black!20, text=black!20}}

\newcommand{\eM}     {\mbox{$\epsilon$-machine}}
\newcommand{\eMs}    {\mbox{$\epsilon$-machines}}
\newcommand{\EM}     {\mbox{$\epsilon$-Machine}}

\newcommand{\MeasAlphabet}  {\mathcal{A}}
\newcommand{\MeasSymbol}   { {X} }
\newcommand{\meassymbol}   { {x} }
\newcommand{\MeasSymbols}[2]{ \MeasSymbol_{#1:#2} }
\newcommand{\meassymbols}[2] { \meassymbol_{#1:#2} }

\newcommand{\Past} { \MeasSymbols{-\infty}{0} }
\newcommand{\pastarrow} { \smash{\overleftarrow {\meassymbol}} }
\newcommand{\past} { \meassymbols{-\infty}{0} }

\newcommand{\pastprimearrow} { {\pastarrow}^{\prime}}
\newcommand{\FutureArrow}   { \smash{\overrightarrow{\MeasSymbol}} }
\newcommand{\Future} { \MeasSymbols{0}{\infty} }

\newcommand{\CausalState}   { \mathcal{S} }

\newcommand{\causalstate}   { \sigma }
\newcommand{\CausalStateSet}    { \boldsymbol{\CausalState} }

\newcommand{\Prob}      {\Pr} 

\newcommand{\Cmu}       {C_\mu}
\newcommand{\hmu}       {h_\mu}
\newcommand{\EE}        {{\bf E}}

\newcommand{\PC}        {\chi}

\newcommand{\ProcessAlphabet}   {\MeasAlphabet}

\newcommand{\forward}{+}
\newcommand{\reverse}{-}
\newcommand{\forwardreverse}{\pm} 

\newcommand{\FutureCausalState} { {\CausalState}^{\forward} }

\newcommand{\PastCausalState}   { {\CausalState}^{\reverse} }

\newcommand{\lastindex}[2]{
  \edef\tempa{0}
  \edef\tempb{#2}
  \ifx\tempa\tempb
    \edef\tempc{#1}
  \else
    \edef\tempa{0}
    \edef\tempb{#1}
    \ifx\tempa\tempb
      \edef\tempc{#2}
    \else
      \edef\tempc{#1+#2}
    \fi
  \fi
  \tempc
}

\newcommand{\COrder}{k_{\PC}}

\newcommand{\MOrder}{R}

\newcommand{\CSjoint}[1][,]{
   \edef\tempa{:}
   \edef\tempb{#1}
   \ifx\tempa\tempb
      \ensuremath{\FutureCausalState\!#1\PastCausalState}
   \else
      \ensuremath{\FutureCausalState#1\PastCausalState}
   \fi
}

\newif\ifpm
\edef\tempa{\forwardreverse}
\edef\tempb{\pm}
\ifx\tempa\tempb
   \pmfalse
\else
   \pmtrue
\fi

\begin{document}

\title{Occam's Quantum Strop:\\
Synchronizing and Compressing Classical Cryptic Processes\\
via a Quantum Channel}



\author{John R. Mahoney}
\email{jrmahoney@ucdavis.edu}
\author{Cina Aghamohammadi}
\email{caghamohammadi@ucdavis.edu}
\author{James P. Crutchfield}
\email{chaos@ucdavis.edu}
\affiliation{Complexity Sciences Center and Department of Physics,
University of California at Davis, One Shields Avenue, Davis, CA 95616}

\date{\today}
\bibliographystyle{unsrt}


\begin{abstract}
A stochastic process's statistical complexity stands out as a fundamental
property: the minimum information required to synchronize one process generator
to another. How much information is required, though, when synchronizing over a
quantum channel? Recent work demonstrated that representing causal similarity
as quantum state-indistinguishability provides a quantum advantage. We
generalize this to synchronization and offer a sequence of constructions
that exploit extended causal structures, finding substantial increase
of the quantum advantage. We demonstrate that maximum compression is determined
by the process's cryptic order---a classical, topological property closely
allied to Markov order, itself a measure of historical dependence. We introduce
an efficient algorithm that computes the quantum advantage and close noting
that the advantage comes at a cost---one trades off prediction for generation
complexity.
\end{abstract}


\preprint{Santa Fe Institute Working Paper 15-07-XXX}
\preprint{arxiv.org:1507.XXXXX [XXX]}

\maketitle

%



Discovering and describing correlation and pattern are critical to progress in
the physical sciences. Observing the weather in California last Summer we find
a long series of sunny days interrupted only rarely by rain---a pattern now all
too familiar to residents. Analogously, a one-dimensional spin system in a
magnetic field might have most of its spins ``up'' with just a few
``down''---defects determined by the details of spin coupling and thermal
fluctuations. Though nominally the same pattern, the domains of these systems
span the macroscopic to the microscopic, the multi-layer to the pure. Despite
the gap, can we meaningfully compare these two patterns?

To exist on an equal descriptive footing, they must each be abstracted from their
physical embodiment by, for example, expressing their generating mechanisms via
minimal probabilistic encodings. Measures of unpredictability, memory, and
structure then naturally arise as information-theoretic properties of these
encodings. Indeed, the fundamental interpretation of (Shannon) information is
as a rate of encoding such sequences. This recasts the informational properties
as answers to distinct communication problems. For instance, a process'
structure becomes the problem of two observers, Alice and Bob, synchronizing
their predictions of the process.

However, what if the communication between Alice and Bob is not classical? What
if Alice instead sends qubits to Bob---that is, they synchronize over a quantum
channel? Does this change the communication requirements? More generally, does quantum
communication enhance our understanding of what ``pattern'' is in the first
place? What if the original process is itself quantum? More practically, is the
quantum encoding more compact?

A provocative answer to the last question appeared recently
\cite{Gu12a, Tan14, Gmein11a} suggesting that a quantum representation can compress a
stochastic process beyond its known classical limits \cite{Crut12a}. In the
following, we introduce a new construction for quantum channels that
 improves and broadens that result to any memoryful stochastic
process, is highly computationally efficient, and identifies optimal quantum
compression. Importantly, we draw out the connection between quantum
compressibility and process cryptic order---a purely classical property that
was only recently discovered \cite{Crut08a}. Finally, we discuss the
subtle way in which the quantum framing of pattern and structure differs from
the classical.


\paragraph*{Synchronizing Classical Processes}
To frame these questions precisely, we focus on patterns generated by
discrete-valued, discrete-time stationary stochastic processes. There is a
broad literature that addresses such emergent patterns \cite{Ball99a, Hoyl06a,
Kant06a}. In particular, computational mechanics is a well developed theory of
pattern whose primary construct---the \emph{\eM}---is a process's minimal,
unifilar predictor \cite{Crut12a}. The \eM's \emph{causal states} $\st
\in \CausalStateSet$ are defined by the equivalence relation that groups all
histories $\pastarrow = \past$ that lead to the same prediction of the future $\FutureArrow = \Future$:
\begin{align}
  \pastarrow & \sim \pastprimearrow \iff \Prob(\FutureArrow | \pastarrow) = \Prob(\FutureArrow | \pastprimearrow)
  ~.
\label{eq:PredEqReln}
\end{align}
Helpfully, a process' \eM\ allows one to directly calculate its measures
of unpredictability, memory, and structure.

For example, the most basic question about unpredictability is, how much
uncertainty about the next future observation remains given complete knowledge
of the infinite past? This is measured by the well-known Shannon
\emph{entropy rate} $\hmu$ \cite{Crut01a, Shan48a, Kolm58,Sina59}:
\begin{align*}
  \hmu = \lim_{L \to \infty} \H(\MS_{L} | \MSs{0}{L})
  ~,
\end{align*}
where $\MSs{0}{L}$ denotes the block of symbols of length $L$, and $\H = - \sum {p_i
\log p_i}$ is the Shannon entropy (in bits using log base 2) of the probability distribution $\{p_i\}$
\cite{Cove06a}. A process's \eM\ allows us to directly calculate this in
closed form as the state-averaged branching uncertainty:
\begin{align*}
  \hmu = \sum \limits_{\st_i \in \CausalStateSet} \pi_i \H(\MS_0 | \St_0 = \st_i)
  ~,
\end{align*}
where $\pi$ denotes the stationary distribution over the causal states.
This form is possible due to \eM's \emph{unifilarity}: in each state $\st$, each symbol $\ms$ leads to at most one successor state $\st'$.

One can ask the complementary question, given knowledge of the infinite past,
how much can we reduce our uncertainty about the future? This quantity is the
mutual information between the past and future and is known the \emph{excess
entropy} \cite[and citations therein]{Crut01a}:
\begin{align*}
\EE & = \I[\Past : \Future] ~.
\end{align*}
It is the total amount of future information \emph{predictable} from the past.
Using the \eM\ we can directly calculate it also:
\begin{align*}
\EE & = \I[\FutureCausalState : \PastCausalState]
  ~,
\end{align*}
where $\FutureCausalState$ and $\PastCausalState$ are the forward (predictive)
and reverse (retrodictive) causal states, respectively \cite{Crut08a}.
This suggests we think of any process as channel that communicates the past to
the future through the present. In this view $\EE$ is the information
transmission rate through the present ``channel''. The excess entropy has been
applied to capture the total predictable information in such diverse systems as
Ising spin models \cite{Crut97a}, diffusion in nonlinear potentials
\cite{Marz14a}, neural spike trains \cite{Stev96a,Bial00a,Marz14e}, and human
language \cite{Debo11a}.

What memory is necessary to \emph{implement} predicting $\EE$ bits of the
future given the past? Said differently, what resources are required to
instantiate this putative channel? Most basically, this is simply the
historical information the process remembers and stores in the present. The
minimum necessary such information is that stored in the causal states, the
\emph{statistical complexity} \cite{Crut12a}:
\begin{align*}
  \Cmu =  \H(\CausalState) = -\sum \limits_i \pi_i \log \pi_i
  ~.
\end{align*}
Importantly, it is lower-bounded by the excess entropy:
\begin{align*}
  \EE \leq \Cmu
  ~,
\end{align*}
where $\pi_i$ denotes the stationary distribution over causal states.

What do these quantities tell us? Perhaps the most surprising observation is
that there is a large class of \emph{cryptic processes} for which $\EE \ll
\Cmu$ \cite{Crut08a}. The structural mechanism behind this difference
is characterized by the \emph{cryptic order}: the minimum $k$ for which
$\H[\St_k | \Future] = 0$. A related and more familiar property is the
\emph{Markov order}: the smallest $R$ for which $\H[\St_R | \MSs{0}{R}] = 0$.
Markov order reflects a process's historical dependence. These orders are
independent apart from the fact that $k \leq R$ \cite{Maho09a, Maho11a}. It is
worth pointing out that the equality $\EE = \Cmu$ is obtained exactly for
cryptic order $k=0$ and, furthermore, that this corresponds with
\emph{counifilarity}---for each state $\st'$ and each symbol $x$, there is at
most one prior state $\st$ that leads to $\st'$ on a transition generating $x$ \cite{Maho11a}.

These properties play a key role in the following communication scenario
where we have a given process's \eM\ in hand. Alice and Bob each have a copy.
Since she has been following the process for some time, using her \eM\ Alice
knows that the process is currently in state $\st_j$, say. From this knowledge,
she can use her \eM\ to make the optimal probabilistic prediction
$\Pr(\MSs{0}{L}|\st_j)$ about the process' future (and do so over arbitrarily
long horizons $L$). While Bob is able to produce all such predictions from each
of his \eM's states, he does not know which particular state is currently
relevant to Alice. We say that Bob and Alice are \emph{unsynchronized}.

To communicate the relevant state to Bob, Alice must send at least $\Cmu$ bits
of information. More precisely, to communicate this information for an ensemble
(size $N \to \infty$) of \eMs, she may, by the Shannon noiseless coding theorem
\cite{Cove06a}, send $N \Cmu$ bits. Under this interpretation, $\Cmu$ is a
fundamental measure of a process's structure in that it characterizes not only
the correlation between past and future, but also the \emph{mechanism} of
prediction. In the scenario with Alice and Bob, $\Cmu$ is seen as the
communication cost to synchronize. We can also imagine Alice using this channel
to communicate with her future self. In this light, $\Cmu$ is understood as a
fundamental measure of a process' internal memory.

\section*{Results}

\subsection{Quantum Synchronization}

\newcommand{\qM} {q-machine}
\newcommand{\qMs} {q-machines}
\newcommand{\qML} {q-machine $M(L)$}
\newcommand{\qMOne} {q-machine $M(1)$}
\newcommand{\qMThree} {q-machine $M(3)$}
\newcommand{\qMThousand} {q-machine $M(1000)$}
\newcommand{\qMk} {q-machine $M(k)$}

What if Alice can send \emph{qubits} to Bob? Consider a communication protocol
in which Alice encodes the causal state in a quantum state that is sent to Bob.
Bob then extracts the information through measurement of this
quantum state. Their communication is implemented via a quantum object---the
\emph{\qM}---that simulates the original stochastic process. It sports a
single parameter that sets the horizon-length $L$ of future words incorporated in the
quantum-state superpositions it employs. We monitor the \qM\ protocol's
efficacy by comparing the quantum-state information transmission rate to the
classical causal-state rate ($\Cmu$).

The \qML\ consists of a set $\{ \ket{\eta_k(L)} \}$ of pure \emph{signal
states} that are in one-to-one correspondence with the classical causal
states $\st_k \in \CausalStateSet$. Each signal state $\ket{\eta_k(L)}$ 
encodes the set of length-$L$ words that may follow $\st_k$, as
well as each corresponding conditional probability used for prediction from
$\st_k$. Fixing $L$, we construct quantum states of the form:
\begin{align*}
\ket{\eta_j(L)} \equiv
  \sum \limits_{w^L \in |\Abet|^L}
  \sum \limits_{\st_k \in \CausalStateSet}
  {\sqrt{\Prob(w^L, \st_k | \st_j)}
  ~ \ket{w^L} \ket{\st_k}}
  ~,
\end{align*}
where $w^L$ denotes a length-$L$ word and $\Prob(w^L,\st_k | \st_j) =
\Prob(X_{0:L} = w^L,\St_L = \st_k | \St_0 = \st_j)$. Due to \eM\ unifilarity, a
word $w^L$ following a causal state $\st_j$ leads to only one subsequent causal
state. Thus, $\Prob(w^L, \st_k | \st_j) = \Prob(w^L | \st_j)$.  The resulting
Hilbert space is the product $\mathcal{H}_w \otimes \mathcal{H}_{\sigma}$.
Factor space $\mathcal{H}_{\sigma}$ is of size $|\mathcal{S}|$, the number of
classical causal states, with basis elements $\ket{\causalstate_k}$. Factor
space $\mathcal{H}_{w}$ is of size $|\mathcal{A}|^L$, the number of length-$L$
words, with basis elements $\ket{w^L} = \ket{x_0} \cdots \ket{x_{L-1}}$.

Note that the $L = 1$ \qMOne\ is equivalent to the construction introduced in
Ref.~\cite{Gu12a}. Additionally, insight about the \qM\ can be gained through
its connection with the classical concatenation machine defined in
Ref.~\cite{Kar11a}; the \qML\ is equivalent to the \qMOne\ derived from the $L$th
concatenation machine.

Having specified the Hilbert state space, we now describe how the \qM\
produces symbol sequences. Given one of the pure quantum signal states, we
perform a projective measurement in the $\mathcal{H}_w$ basis. This results in
a symbol string $w^L = x_0, \ldots, x_{L-1}$, which we take as the next $L$
symbols in the generated process. Since the \eM\ is unifilar, the quantum
conditional state must be in some basis state $\ket{\st_k}$ of
$\mathcal{H}_{\sigma}$. Subsequent measurement in this basis then indicates the
corresponding classical causal state with no uncertainty.

Observe that the probability of a word $w^L$ given quantum state $\ket{\eta_k}$
is equal to the probability of that word given the analogous classical state
$\causalstate_k$. Also, the classical knowledge of the subsequent corresponding
causal state can be used to prepare a subsequent quantum state for continued
symbol generation. Thus, the \qM\ generates the desired stochastic process and
is, in this sense, equivalent to the classical \eM.

Focus now on the \qM's initial quantum state:
\begin{align*}
\rho(L) = \sum \limits_i {p_i \ket{\eta_i(L)} \bra{\eta_i(L)}}
  ~.
\end{align*}
We see this mixed quantum state is composed of pure signal states combined
according to the probabilities of each being prepared by Alice (or being
realized by the original process that she observes). These are simply the
probabilities of each corresponding classical causal state, which we take to be
the stationary distribution: $p_i = \pi_i$. In short, quantum state $\rho(L)$
is what Alice must transmit to Bob for him to successfully synchronize.
Later, we revisit this scenario to discuss the tradeoffs associated with the \qM\ representation.

If Alice sends a large number $N$ of these states, she may, according to the
quantum noiseless coding theorem \cite{Schu95}, compress this message into $N
S(\rho(L))$ qubits, where $S$ is the von Neumann entropy $S(\rho) = \mathrm{tr}
(\rho \log(\rho))$. Due to its parallel with $\Cmu$, and for convenience, we
define the function:
\begin{align*}
C_q(L) \equiv S(\rho(L))
  ~.
\end{align*}
Recall that, classically, Alice must send $N \Cmu$ bits. To the extent that
$N C_q (L)$ is smaller, the quantum protocol will be more efficient.



\subsection{Example Processes: $C_q(L)$}

Let's now draw out specific consequences of using the \qM. We explore protocol
efficiency by calculating $C_q(L)$ for several example processes, each chosen
to illustrate distinct properties: \qM\ affords a quantum advantage, further
compression can be found at longer horizons $L$, and the compression rate is
minimized at the horizon length $k$---the cryptic order of the classical process
\cite{Maho11a}.

For each example, we examine a process family by sweeping one transition
probability parameter, illustrating $C_q(L)$ and its relation to classical bounds $\Cmu$ and $\EE$.
Additionally, we highlight a single
representative process at one generic transition probability. 
Following these examples, we turn to discuss more general properties of \qM\ compression
that apply quite broadly and how the results alter our notion of quantum
structural complexity.

\subsubsection{Biased Coins Process}

\begin{figure}
\includegraphics[width=0.6\linewidth]{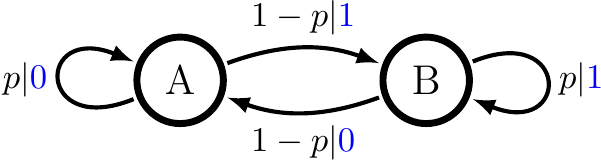}
%
\includegraphics[width=\linewidth]{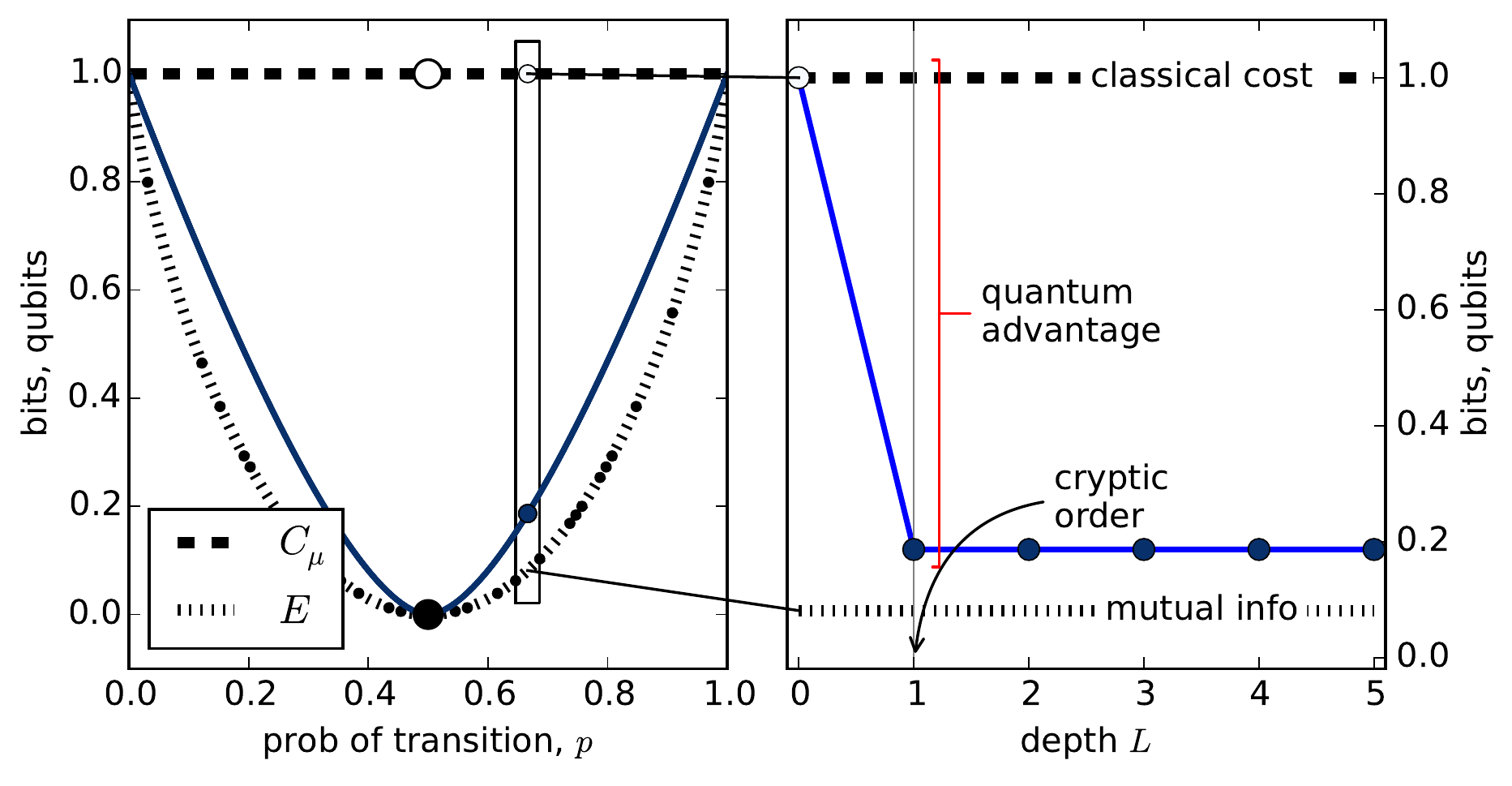}
\caption{Biased Coins Process:
  (top) \EM. Edges are conditional probabilities. For example, self-loop on 
  state $A$ $p|\textcolor{blue}{0}$ indicates $\Pr(\textcolor{blue}{0} | A) = p$.
  (left) Statistical complexity $\Cmu$, quantum state entropy $C_q(L)$, and
  excess entropy $\EE$ as a function of $A$'s self-loop probability $p \in
  [0,1]$. $C_q(1)$ (dark blue) lies between $\Cmu$ and $\EE$ (bits),
  except for extreme parameters and the center ($p = 1/2$).
  (right) For $p=0.666$, $C_q(L)$ decreases from $L=0$ to $L=1$ and is then
  constant; the process is maximally compressed at $L=1$, its cryptic order
  $k = 1$. This yields substantial compression: $C_q(1) \ll \Cmu$.
  }
\label{fig:perturbed_coin_vne}
\end{figure}

The Biased Coins Process provides a first, simple case that realizes a
nontrivial quantum state entropy \cite{Gu12a}. There are two biased coins,
named $A$ and $B$. The first generates $0$ with probability $p$; the second,
$0$ with probability $1-p$. A coin is picked and flipped, generating some output
$0$ or $1$. With probability $1-p$ the other coin is used next. Its two
causal-state \eM\ is shown in Fig.~\ref{fig:perturbed_coin_vne}(top).

Consider $p \approx 1/2$. The generated sequence is close to that of a fair
coin. And, starting with coin $A$ or $B$ makes little difference to the
future.  There is little to predict about future sequences. This intuition is
quantified by the predictable information $\EE \approx 0$, when $p$ is near
$1/2$. See Fig.~\ref{fig:perturbed_coin_vne}(left).

In contrast, since the causal states have equal probability, $\Cmu = 1$ bit
independent of parameter $p$. (All informations are quoted in log base $2$.)
The gap between $\Cmu$ and $\EE$ presents an opportunity for large quantum
improvement. This is because there is always \emph{some}, albeit very little,
predictive advantage to remembering whether the last symbol was $0$ or $1$.
Retaining this advantage, however small, requires the use of an entire
(classical) bit. It is only at the exact value $p=1/2$ where the two causal
states merge, this advantage disappears, and the process becomes memoryless
(IID). This is reflected in the discontinuity of $\Cmu$ as $p \to 1/2$, which
is sometimes misinterpreted as a deficiency of $\Cmu$. Contrariwise, this
feature follows naturally from the equivalence relation and is a signature of
symmetry.

Now, let's consider these complexities in the quantum setting where we monitor
communication costs using $C_q(L)$. To understand its behavior, we first write
down the \qM's states. For $L=0$, we have the trivial $\ket{\eta^0_A} =
\ket{A}$ and $\ket{\eta^0_B} = \ket{B}$. For $L=1$, we have $\ket{\eta^1_A} =
\sqrt{1-p} \ket{0} \ket{A} + \sqrt{p} \ket{1} \ket{B}$ and $\ket{\eta^1_B} =
\sqrt{p} \ket{0} \ket{A} + \sqrt{1-p} \ket{1} \ket{B}$. The von Neumann entropy
of the former is simply the Shannon information of the signal state
distribution; that is, $C_q(0) = \Cmu$. In the latter, however, the two quantum
states have a nonzero overlap (inner product). This implies that the von Neumann entropy is
smaller than the Shannon entropy $C_q(1) < \Cmu = C_q(0)$. (See Ref.~\cite{Niel11} Thm.~11.10.) Also, making use of the Holevo bound, we see that
$\EE \leq C_q(1)$ \cite{Hole73, Gu12a}. These bounds are maintained for all
$L$: $\EE \leq C_q(L) \leq \Cmu$. This follows by considering the \qMOne\ of
the $L$th classical concatenation.

(Note that for $p \in \{0, 1/2, 1\}$ these quantities are all equal and equal to
zero.  This comes from the simplification of process topology caused by state
merging dictated by the predictive equivalence relation, Eq.
(\ref{eq:PredEqReln}).)

How do costs change with sequence length $L$? To see this
Fig.~\ref{fig:perturbed_coin_vne}(right) expands the left view, but for a
single value of $p$. As expected, $C_q(L)$ decreases from $L=0$ to $L=1$.
However, it then remains constant for all $L \ge 1$. There is no additional
quantum state-compression afforded by expanding the \qM\ to use longer horizons.

The Biased Coins Process has been analyzed earlier using a construction
equivalent to an $L = 1$ \qM\ \cite{Gu12a}, similarly finding that that the number of
required qubits falls between $\EE$ and $\Cmu$. The explanation there for this
compression ($C_q(1) < \Cmu$) was lack of counifilarity in the process' \eM.
More specifically, Ref.~\cite{Gu12a} showed that $\EE = C_q = \Cmu$ if and only
if the \eM\ is counifilar, and $\EE < C_q < \Cmu$ otherwise. The Biased Coins
Process is easily seen to be noncounifilar and so the inequality follows.

This previous analysis happens to be sufficient for the Biased Coins Process,
since $C_q(L)$ does not decrease beyond $L=1$. Unfortunately, only this single,
two-state process was analyzed when, in fact, 
the space of processes is replete with richly structured behaviors
\cite{Feld08a}. With this in mind and to show the power of the \qM, we step
into deeper water to consider a $7$-state process that is almost periodic with
a random phase-slip.

\begin{figure}
\includegraphics[width=0.8\linewidth]{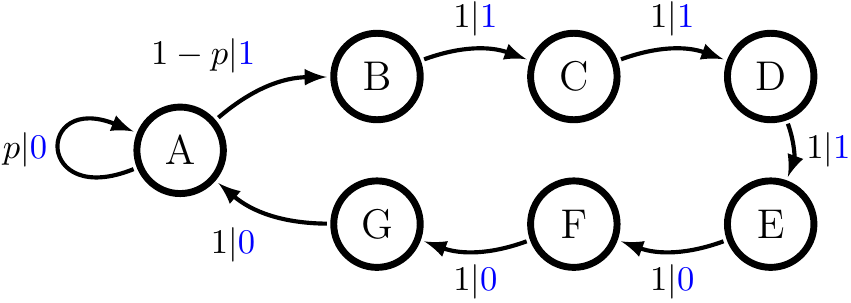}
%
\includegraphics[width=\linewidth]{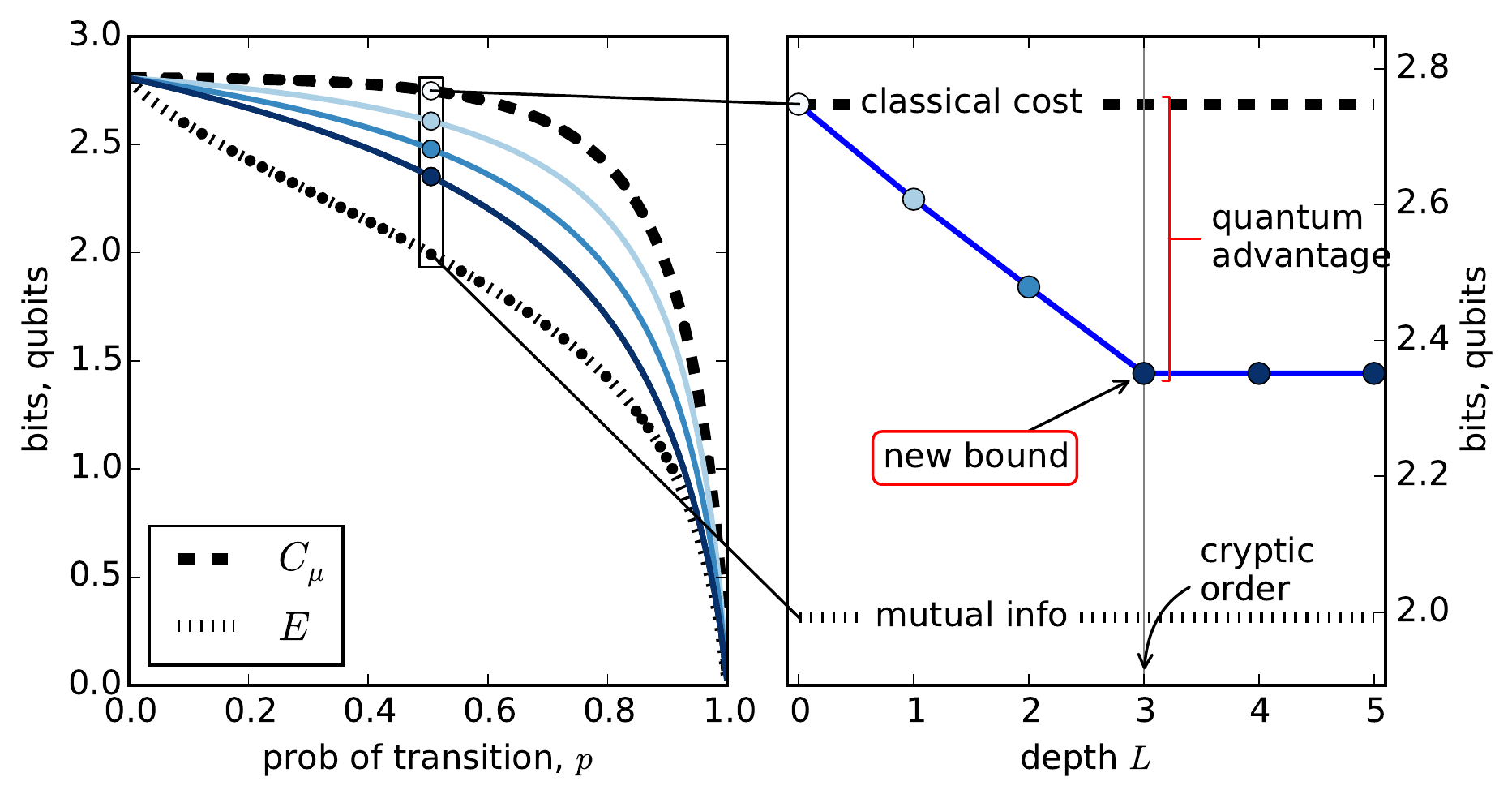}
\caption{$4$-$3$ Golden Mean Process:
  (top) The \eM.
  (left) Statistical complexity $\Cmu$, quantum state entropy $C_q(L)$, and
  excess entropy $\EE$ as a function of $A$'s self-loop probability $p \in
  [0,1]$. $C_q(L)$ is calculated and plotted (light to dark blue) up to $L=5$.
  (right) For $p=0.505$, $C_q(L)$ decreases monotonically until $L=3$---the
  process' cryptic order. The additional compression is
  substantial: $C_q(3) \ll C_q(1)$.
  }
\label{fig:rk_golden_mean_vne}
\end{figure}

\subsubsection{$R$-$k$ Golden Mean Process}

The $R$-$k$ Golden Mean Process is a useful generalization of the Markov
order-$1$ Golden Mean Process that allows for the independent specification of
Markov order $R$ and cryptic order $k$ \cite{Maho09a, Maho11a}.
Figure~\ref{fig:rk_golden_mean_vne}(top) illustrates its \eM. We take $R=4$ and $k=3$.

The calculations in Fig.~\ref{fig:rk_golden_mean_vne}(left) show again that
$C_q(L)$ generically lies between $\EE$ and $\Cmu$, across this family of
processes. In contrast with the previous example, $C_q(L)$ continues to
decrease beyond $L=1$. Figure~\ref{fig:rk_golden_mean_vne}(right) illustrates
that the successive \qMs\ continue to reduce the von Neumann entropy: $\Cmu >
C_q(1) > C_q(2) > C_q(3)$.  However, there is no further improvement beyond a
future-depth of $L = 3$, the cryptic order: $C_q(3) = C_q(L>3)$. It is important to note that the
compression improvements at stages $L = 2$ and $L = 3$ are significant.
Therefore, a length-$1$ quantum representation misses the majority of the
quantum advantage.

To understand these results we need to sort out how quantum compression stems
from noncounifilarity. In short, the latter leads to quantum signal states with
nonzero overlap that allow for super-classical compression. Let's explain
using the current example. There is one noncounifilar state in this process,
state $A$. Both states $A$ and $G$ lead to $A$ on a symbol $1$. Due to this, at
$L=1$, the two \qM\ states:
\begin{align*}
\ket{\eta_A} &= \sqrt{p} \ket{1} \ket{A} + \sqrt{1-p} \ket{0} \ket{B}
  ~\text{and}~ \\
  \ket{\eta_G} &= \ket{1} \ket{A}
\end{align*}
have a nonzero overlap of $\braket{\eta_A | \eta_G} = \sqrt{p}$.
(All other overlaps in the $L = 1$ \qM\ vanish.)
As with the Biased Coins Process, this leads to the inequality $C_q(1) < \Cmu$.

Extending the representation to $L=2$ words, we find three nonorthogonal quantum states:
\begin{align*}
\ket{\eta_A} &= \sqrt{p^2} \ket{11} \ket{A} + \sqrt{p(1-p)} \ket{10} \ket{B} \\
             & \quad\quad\quad + \sqrt{(1-p)} \ket{00} \ket{C} ~,\\
\ket{\eta_F} &= \ket{11} \ket{A} ~,~\text{and}\\
\ket{\eta_G} &= \sqrt{p} \ket{11} \ket{A} + \sqrt{1-p} \ket{10} \ket{B} ~,
\end{align*}
with three nonzero overlaps $\braket{\eta_A | \eta_F} = p$, $\braket{\eta_A | \eta_G} = \sqrt{p}$, and $\braket{\eta_F | \eta_G} = \sqrt{p}$.

Note that the overlap $\braket{\eta_A | \eta_G}$ is unchanged.
This is because the conditional futures are identical once the merger on symbol $1$ has taken place.
That is, the words $11$ and $10$, which contribute to the $L=2$
$\braket{\eta_A | \eta_G}$ overlap, simply derive from the prefix $1$, which was
the source of the overlap at $L=1$. In order to obtain a change
in this or any other overlap, there must be a \emph{new} merger-inducing prefix
(for that state-pair). (See Sec.~\ref{sec:methods} for computational
implications.) Since all quantum amplitudes are positive, each pairwise overlap
is a nondecreasing function of $L$.

At $L=2$ we have two such new mergers: $11$ for $\braket{\eta_A | \eta_F}$ and
$11$ for $\braket{\eta_F | \eta_G}$. This additional increase in pairwise
overlaps leads to a second decrease in the von Neumann entropy. (See
Sec.~\ref{sec:monotonicity} for details.) Then, at $L=3$, we find three new
mergers: $111$ for $\braket{\eta_A | \eta_E}$, $111$ for $\braket{\eta_E |
\eta_F}$, and $111$ for $\braket{\eta_E | \eta_G}$. As before, the pre-existing
mergers simply acquire suffixes and do not change the degree of overlap.

Importantly, we find that at $L=4$ there are no new mergers. That is, any
length-$4$ word that leads to the merging of two states must merge before the
fourth symbol. In general, the length at which the last merger occurs is
equivalent to the cryptic order \cite{Maho11a}. Further, it is known that the
von Neumann entropy is a function of pairwise overlaps of signal states
\cite{Jozsa00a}. Therefore, a lack of new mergers, and thus constant overlaps,
implies that the von Neumann entropy is constant. This demonstrates that
$C_q(L)$ is constant for $L \geq k$, for $k$ the cryptic order.

The $R$-$k$ Golden Mean Process was selected to highlight the unique role of
the cryptic order, by drawing a distinction between it and Markov order. The
result emphasizes the physical significance of the cryptic order. In the
example, it is not until $L=4$ that a naive observer can synchronize to the
causal state; this is shown by the Markov order. For example, the word $000$
induces two states $D$ and $E$. Just one more symbol synchronizes to either $E$
(on $0$) or $F$ (on $1$). Yet recall that synchronization can come about in two
ways. A word may either induce a path merger or a path termination. All
merger-type synchronizations must occur no later than the last termination-type
synchronization. This is equivalently stated: the cryptic order is never
greater than the Markov order \cite{Maho11a}.

In the current example, we observe this termination-type of synchronization on
the symbol following $000$. For instance, $0000$ does not lead to the merger of
paths originating in multiple states. Rather, it eliminates the possibility
that the original state might have been $B$.

It is the final merger-type synchronization at $L=3$ that leads to the final
unique-prefix quantum merger and, thus, to the ultimate minimization of the von
Neumann entropy. So, we see that in the context of the \qM, the most efficient state compression is accomplished at the process's cryptic order. (One could certainly continue beyond the cryptic order, but at best this increases implementation cost with no functional benefit.)

\begin{figure}
\includegraphics[width=0.5\linewidth]{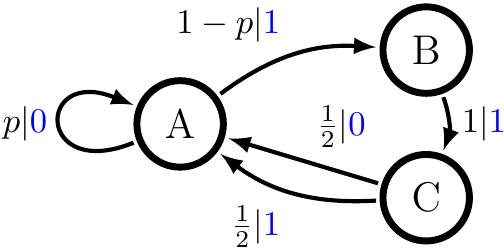}
%
\includegraphics[width=\linewidth]{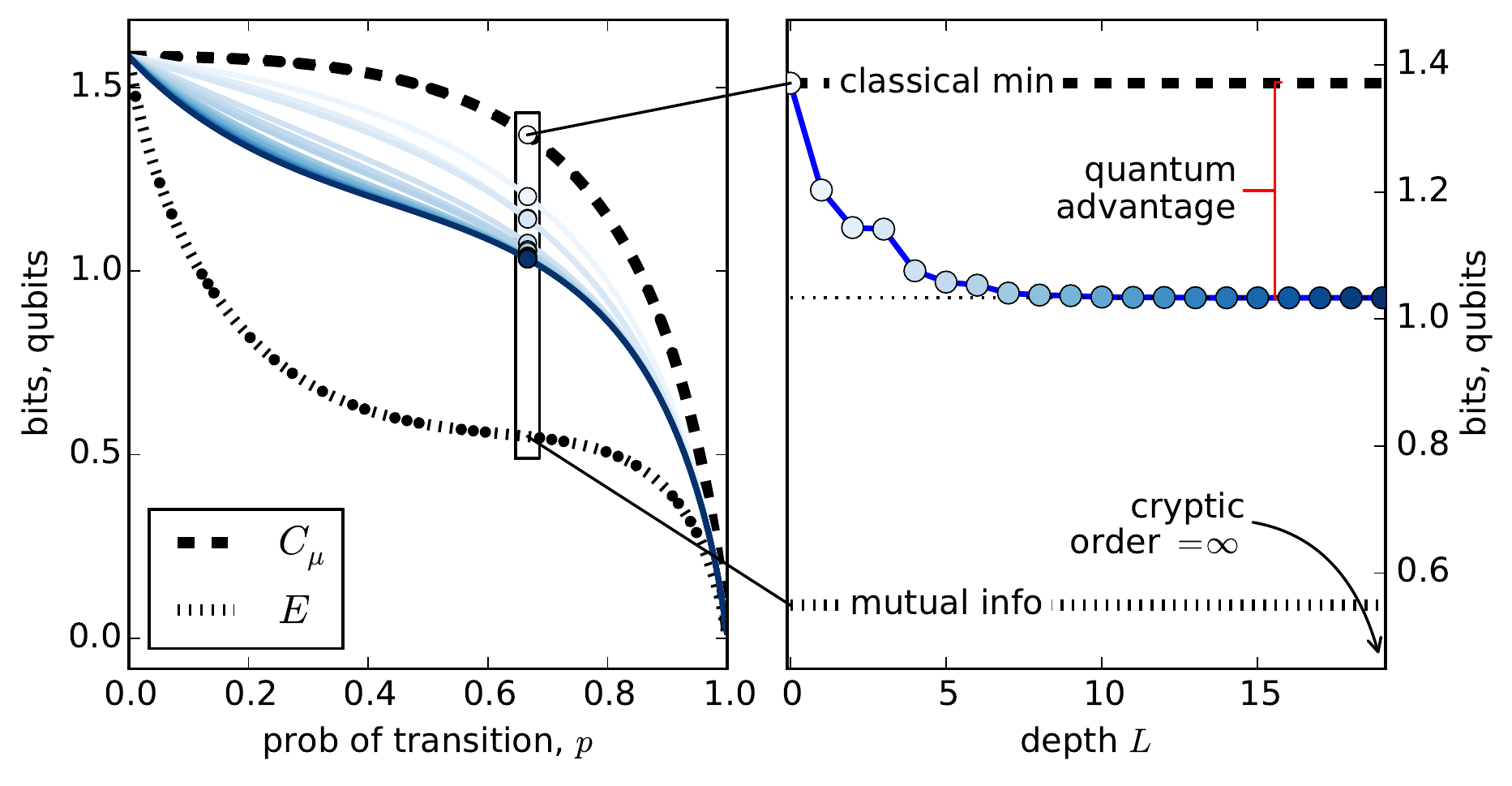}
\caption{Nemo Process:
  (top) Its \eM.
  (left) Statistical complexity $\Cmu$, quantum state entropy $C_q(L)$, and
  excess entropy $\EE$ as a function of $A$'s self-loop probability $p \in
  [0,1]$. $C_q(L)$ is calculated and plotted (light to dark blue) for $L = 0,
  1, .. ,19$.
  (right) For $p=0.666$, $C_q(L)$ decreases monotonically, never reaching
  the limit since the process' cryptic order is infinite.
  The full quantum advantage is realized only in the limit.
  }
\label{fig:nemo_vne}
\end{figure}

\subsubsection{Nemo Process}

To demonstrate the challenges in quantum compressing typical memoryful
stochastic processes, we conclude our set of examples with the seemingly simple
three-state Nemo Process, shown in Fig. \ref{fig:nemo_vne}(top). Despite its
overt simplicity, both Markov and cryptic orders are infinite. As one should
now anticipate, each increase in the length $L$ affords a smaller and smaller
state entropy, yielding the infinite chain of inequalities: $\Cmu \geq C_q(1)
\geq C_q(2) \geq C_q(3) \geq \ldots \geq C_q(\infty) $.
Figure~\ref{fig:nemo_vne}(right) verifies this. This sequence approaches the
asymptotic value $C_q(\infty) \simeq 1.0332$. We also notice that the
convergence of $C_q(L)$ is richer than in the previous processes. For example,
while the sequence monotonically decreases (and at each $p$), it is not convex
in $L$. For instance, the fourth quantum incremental improvement is greater
than the third.

We now turn to discuss the broader theory that underlies the preceding
analyses. We first address the convergence properties of $C_q(L)$, then the
importance of studying the full range of memoryful stochastic processes, and
finally tradeoffs between synchronization, compression, and prediction.


\subsection{$C_q(L)$ Monotonicity}
\label{sec:monotonicity}

It is important to point out that while we observed nonincreasing $C_q(L)$ in
our examples, this does not constitute proof. The latter is nontrivial since
Ref.~\cite{Jozsa00a} showed that each pairwise overlap of signal states can
increase while \emph{also increasing} von Neumann entropy. (This assumes a
constant distribution over signal states.) Furthermore, this phenomenon occurs
with nonzero measure. They also provided a criterion that can exclude this
somewhat nonintuitive behavior. Specifically, if the element-wise ratio matrix
$R$ of two Gram matrices of signal states is a positive operator, then strictly
increasing overlaps imply a decreasing von Neumann entropy. We note, however,
that there exist processes with \eMs\ for which the $R$ matrix is nonpositive.
At the same time, we have found no example of an increasing $C_q(L)$. 

So, while it appears that a new criterion is required to settle this issue, the
preponderance of numerical evidence suggests that $C_q(L)$ is indeed
monotonically decreasing. In particular, we verified $C_q(L)$ monotonicity for
many processes drawn from the topological \eM\ library \cite{John10a}.
Examining $1000$ random samples of two-symbol, $N$-state processes for $2
\leq N \leq 7$ yielded no counterexamples. Thus, failing a proof, the survey
suggests that this is the dominant behavior.

\subsection{Infinite Cryptic Order Dominates}

\begin{figure}
  \centering
  \includegraphics[width=\columnwidth]{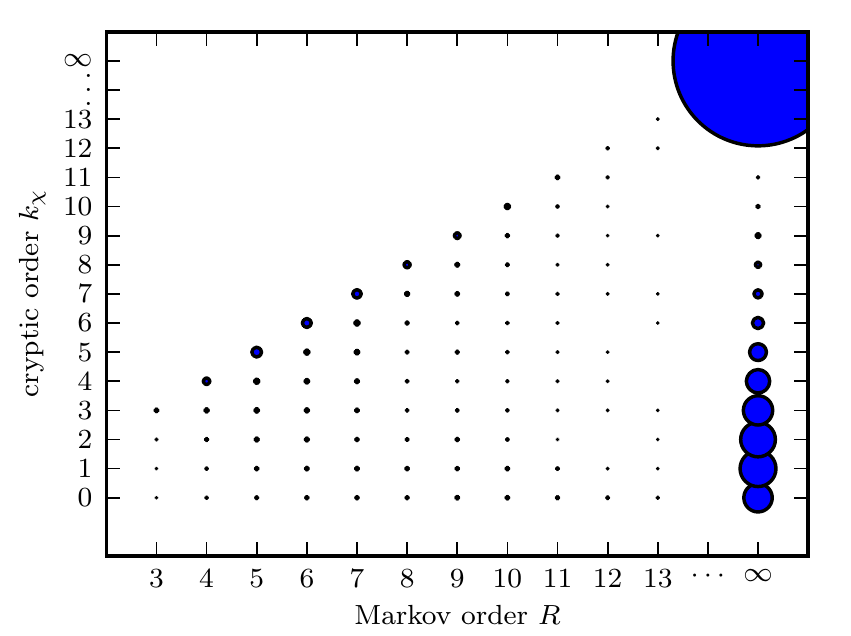}
  \caption{Distribution of Markov order $\MOrder$ and cryptic order $\COrder$
    for all $1,132,613$ six-state, binary-alphabet, exactly-synchronizing \eMs.
    Marker size is proportional to the number of \eMs\ within this class at the
    same ($\MOrder$, $\COrder$).
	(Reprinted with permission from Ref. \cite{Jame10a}.)
	}
\label{fig:rvsk}
\end{figure}
 
The Biased Coins Process, being cryptic order $k=1$, is atypical. Previous
exhaustive surveys demonstrated the ubiquity of infinite Markov and cryptic
orders within process space. For example, Fig.~\ref{fig:rvsk} shows the
distribution of different Markov and cryptic orders for processes generated by
six-state, binary-alphabet, exactly-synchronizing \eMs\ \cite{Jame10a}. The
overwhelming majority have infinite Markov and cryptic orders. Furthermore,
among those with finite cryptic order, orders zero and one are not common. Such
surveys in combination with the apparent monotonic decrease of $C_q(L)$ confirm that, when it comes to general claims
about compressibility and complexity, it is advantageous to extend analyses to
long sequence lengths.

\subsection{Prediction-Compression Trade Off}


\begin{figure}
\includegraphics[width=\linewidth]{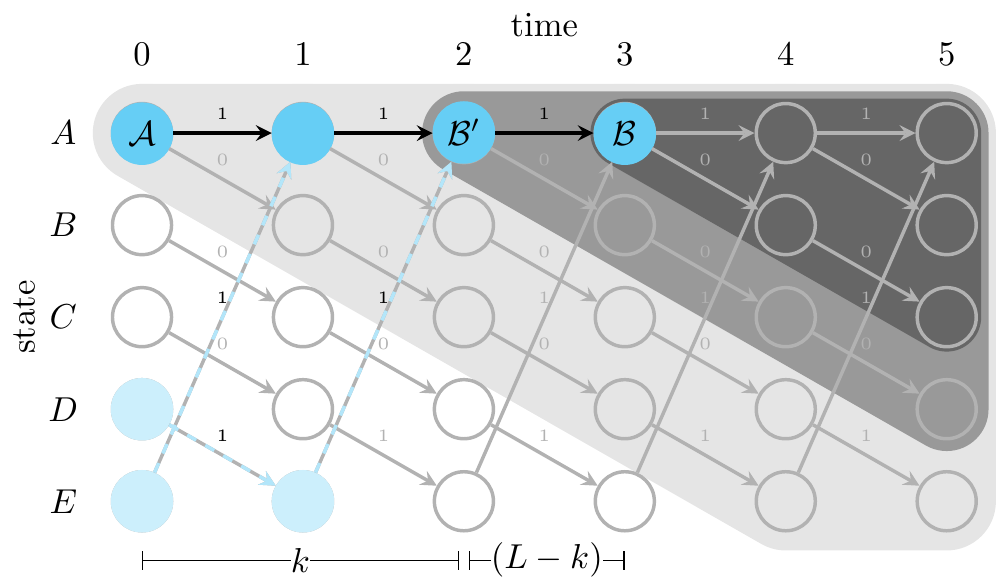}
\caption{Trading prediction for quantum compression: $\mathcal{A}$ is Alice's
  state of predictive knowledge. $\mathcal{B}$ is that for Bob, except when he
  uses the process' \eM\ to refine it. In which case, his predictive knowledge
  becomes that in $\mathcal{B}'$, which can occur at a time no earlier
  than that determined by the cryptic order $k$.
  }
\label{fig:tradeoff}
\end{figure}

Let's return to Alice and Bob in their attempt to synchronize on a given
stochastic process to explore somewhat subtle trade-offs in compressibility,
prediction, and complexity. Figure~\ref{fig:tradeoff} illustrates the
difference in their ability to generate probabilistic predictions about the
future given the historical data. There, Alice is in causal state $A$
(signified by $\mathcal{A}$ for Alice). Her prediction ``cone'' is depicted in
light gray. It depicts the span over which she can generate probabilistic
predictions conditioned on the current causal state ($A$). She chooses to map
this classical causal state to a $L = 3$ \qM\ state and send it to Bob.
(Whether this is part of an ensemble of other such states or not affects the
rate of qubit transmission, but not the following argument.) It is important to
understand that Bob cannot actually determine the corresponding causal state
(at time $t=0$). He can, however, make a measurement that results in some
symbol sequence of length $3$ followed by a definite (classical) causal state.
In the figure, he generates the sequence $111$ followed by causal state $A$ at
time $t=3$. This is shown by the blue state-path ending in $\mathcal{B}$ for
Bob. Now Bob is in position to generate corresponding \emph{conditional}
predictions---$\mathcal{B}$'s future cone $\Pr(\Future|\mathcal{B})$. As the
figure shows, this cone is only a subprediction of Alice's. That is, it is
equivalent to Alice's prediction conditioned on her observation of $111$ or any
other word leading to the same state.

Now, what \emph{can} Bob say about times $t = 0,1,2$? The light blue states and
edges in the figure show the alternate paths that could have also lead to his
measurement of the sequence $111$ and state $A$. For instance, Bob can only say
that Alice might have been in causal states $A$, $D$, or $E$ at time $t=0$. In
short, the quantum representation led to his uncertainty about the initial
state sequence and, in particular, Alice's prediction. All together, we see
that the quantum representation gains compressibility at the expense of Bob's
predictive power.

What if Alice does not bother to compute $k$ and, wanting to make good use of
quantum compressibility, uses an $L = 1000$  \qM? Does this necessarily
translate into Bob's uncertainty in the first $1000$ states and, therefore,
only a highly conditional prediction? In our example, Alice was not quite so
enthusiastic and settled for the $L = 3$ \qM. We see that Bob can use his
current state $A$ at $t=3$ and knowledge of the word that led to it to infer
that the state at $t=2$ must have been $A$. The figure denotes his knowledge of
this state by $\mathcal{B}'$. For other words he may be able to trace farther
back. (For instance, $000$ can be traced back from $D$ at $t=3$ all the way to
$A$ at $t=0$.) The situation chosen in the figure illustrates the worst-case
scenario for this process where he is able to trace back and discover all but
the first 2 states. The worst-case scenario defines the cryptic order $k$, in
this case $k=2$. After this tracing back, Bob is then able to make the improved
statement, ``If Alice observes symbols $11$, then her conditional prediction
will be $\Prob(\Future | A)$''. This means that Alice and Bob cannot suffer
through \emph{overcoding}---using an $L$ in excess of $k$.




Finally, one feature that is unaffected by such manipulations is the ability of
Alice and Bob to \emph{generate} a single future instance drawn from the
distribution $\Prob(\Future | A)$. This helps to emphasize that generation is
distinct from prediction. Note that this is true for the
\qML\ at any length.

\section*{Methods}
\label{sec:methods}

Let's explain computing $C_q(L)$. First, note that the size of the
\qML\ Hilbert space grows as $L^{|A|}$ ($L^{2|A|}$ for the density operators). That
is, computing $C_q(L=20)$ for the Nemo Process involves finding eigenvalues of
a matrix with $10^{12}$ elements. Granted, these matrices are often sparse, but
the number of components in each signal state still grows exponentially with
the topological entropy rate of the process. This alone would drive
computations for even moderately complex processes (described by moderate-sized
\eMs) beyond the access of contemporary computers.

Recall though that there are, at any $L$, still only $|S|$ quantum signal
states to consider. Therefore, the embedding of this constant-sized subspace
wastes an exponential amount of the embedding space. We desire a computation of
$C_q(L)$ that is independent of the diverging embedding dimension.

Another source of difficulty is the exponentially increasing number of words
with $L$. However, we only need to consider a small subset of these words. Once
a merger has occurred between states $\ket{\eta_i}$ and $\ket{\eta_j}$ on word
$w$, subsequent symbols, while maintaining that merger, do not add to the
corresponding overlap. That is, the contribution to the overlap $\braket{\eta_i
| \eta_j}$ by all words with prefix $w$ is complete.

To take advantage of these two opportunities for reduction, we compute $C_q(L)$ in the following manner.

\begin{figure}
\includegraphics[width=0.6\linewidth]{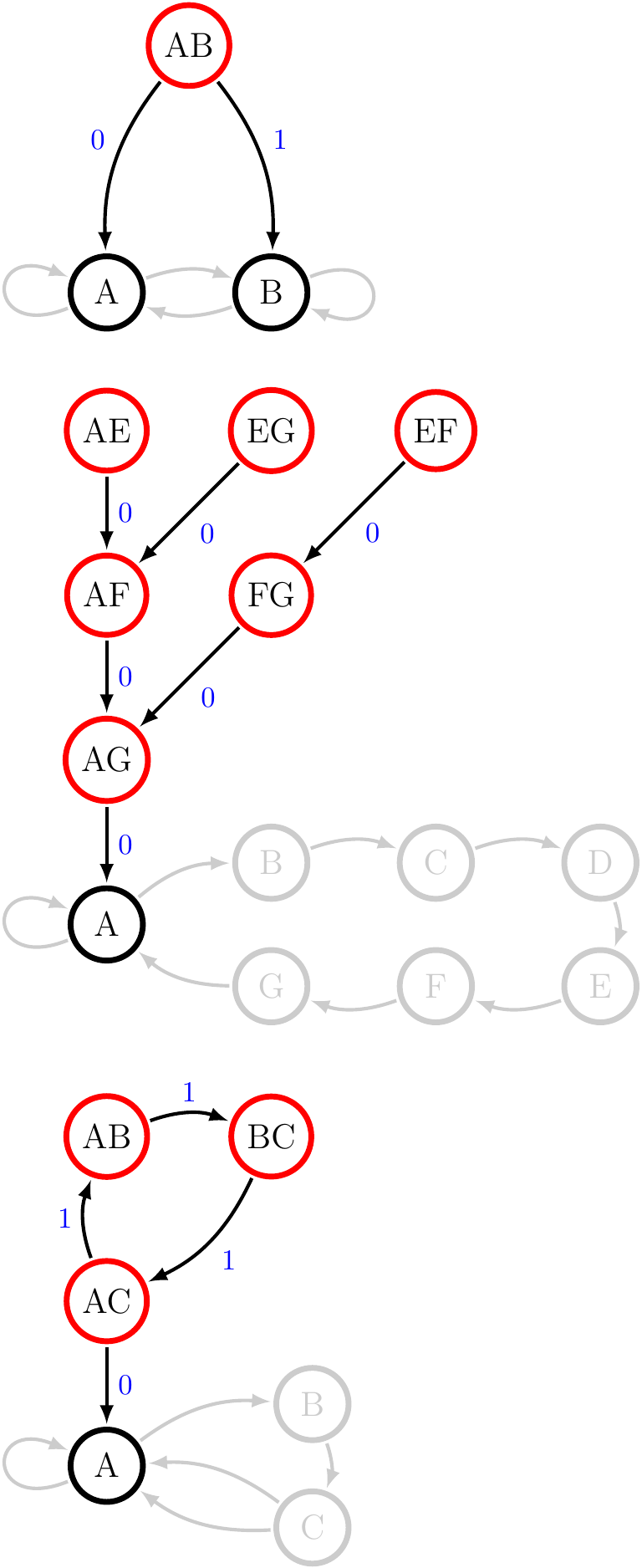}

\caption{Pairwise-merger machines for our three example processes. Pair-states (red) lead to each other or enter the \eM\ at a noncounifilar state. For example, in the R-k Golden Mean (middle), the two pair-states $AF$ and $FG$ both lead to pair-state $AG$ on 0. Then pair-state $AG$ leads to state $A$, the only noncounifilar state in this \eM.}
\label{fig:PMM}
\end{figure}

First, we construct the ``pairwise-merger machine'' (PMM) from the \eM. The
states of the PMM are unordered pairs of causal states. A pair-state
$(\causalstate_i, \causalstate_j)$ leads to $(\causalstate_m, \causalstate_n)$
on symbol $x$ if $\causalstate_i$ leads to $\causalstate_m$ on $x$ and
$\causalstate_j$ leads to $\causalstate_m$ on $x$. (Pairs are unordered, so we
consider $m \leftrightarrow n$ as well.) If both components in a pair-state
lead to the same causal state, then this represents a merger. Of course, these
mergers from pair-states occur only when entering noncounifilar states of the
\eM. If either component state forbids subsequent emission of symbol $x$, then
that edge is omitted. The PMMs for the three example processes are shown in
Fig.~\ref{fig:PMM}.

Now, making use of the PMM, we begin at each noncounifilar state and proceed
backward through the pair-state transient structure. At each horizon-length, we
record the pair-states visited and with what probabilities. This allows
computing each increment to each overlap. Importantly, by moving \emph{up} the
transient structure, we avoid keeping track of any further novel overlaps; they
are all ``behind us''. Additionally, the finite number of pair-states gives us
a finite structure through which to move; when the end of a branch is reached,
its contributions cease. It is worth noting that this pair-state transient
structure may contain cycles (as it does for the Nemo Process). In that case,
the algorithm is non-halting, but it is clear that contributions generated
within a cycle decrease exponentially.

All of this serves to yield the set of overlaps at each length.
We then use a Gram-Schmidt-like procedure to produce a set of $|\mathcal{S}|$ vectors in $\mathbb{R}^{|\mathcal{S}|+}$ (the positive hyperoctant) having the desired set of overlaps.

Weighting these real, positive vectors with the stationary distribution yields
a real, positive-element representation of the density operator restricted to
the subspace spanned by the signal states. At this point, computing $C_q(L)$
reduces to finding eigenvalues of an $|\mathcal{S}| \times |\mathcal{S}|$
matrix.

For example, consider the Nemo Process. The sequence of overlap
\emph{increments} for $L = [0,1,2,3,4,5,6,7,8,\ldots]$, for $\braket{\eta_0 |
\eta_1}$, $\braket{\eta_1 | \eta_2}$, $\braket{\eta_2 | \eta_0}$ respectively,
is given by:
\begin{align*}
\frac{\sqrt{p(1-p)}}{2}
  & \times [0, 0, 0, a^0, a^0, a^0, a^1, a^1, a^1,\ldots] ~,\\
\frac{\sqrt{p}}{2}
  & \times [0, 0, a^0, a^0, a^0, a^1, a^1, a^1,a^2,\ldots] ~, ~\text{and}~\\
\sqrt{\frac{p}{2}}
  & \times [0, a^0, a^0, a^0, a^1, a^1, a^1,a^2, a^2,\ldots]
  ~,
\end{align*}
where $a = (1-p)/2$.


And, the asymptotic cumulative overlaps are given by:
\begin{align*}
\braket{\eta_0 | \eta_1} &= \frac{\sqrt{p(1-p)}}{1+p} ~,\\
\braket{\eta_1 | \eta_2} &= \frac{\sqrt{p}}{1+p} ~, ~\text{and}\\
\braket{\eta_2 | \eta_0} &= \frac{\sqrt{2p}}{1+p}
  ~.
\end{align*}

From this, we computed the restricted density matrix and, hence, its $L \to
\infty$ entropy $C_q(\infty) \simeq 1.0332$, as illustrated in Fig.~\ref{fig:nemo_vne}. The density matrix and eigenvalue forms are long and not
particularly illuminating, and so we do not quote them here. A sequel details a
yet more efficient analytic technique based on holomorphic functions of the
internal-state Markov chain of a related quantum transient structure.

\section*{Discussion}

Recalling our original motivation, we return to the concept of \emph{pattern};
in particular, its representation and characterization. We showed that, to
stand as a canonical form, a process' quantum representation should encode,
explicitly in its states, process correlations over a sufficiently long
horizon-length. In demonstrating this, our examples and analyses found that the
\qM\ generally offers a more efficient quantum representation than the
alternative previously introduced \cite{Gu12a}.

Interestingly, the length scale at which our construction's compression
saturates is the cryptic order, a recently introduced measure of causal-state
merging and synchronization for classical stochastic processes. Cryptic order, in contrast
to counifilarity, makes a finer division of process space, suggesting that it
is a more appropriate explanation for super-classical compression. We also
developed efficient algorithms to compute this ultimate quantum
compressibility. Their computational efficiency is especially important for
large or infinite cryptic orders, which are known to dominate process space.

We cannot yet establish the minimality of our construction with respect to all
alternatives. For example, more general quantum hidden Markov models (QHMMs)
may yield a greater advantage \cite{Gmein11a}. Proving minimality among QHMMs
is of great interest on its own, too, as it should lead to a canonical quantum
representation of classical stochastic processes. States in such QHMMs might
then be appropriately named ``quantum causal states''.




And, what is the meaning of the remaining gap between $C_q(k)$ and $\EE$? In
the case that $C_q(k)$ is in fact a minimum, this difference should represent a
quantum analog of the classical crypticity. Physically, since the latter
is connected with information thermodynamic efficiency \cite{Elli11a}, it would
then control the efficiency for quantum thermodynamic processes.



Let's close by outlining future impacts of these results. Most generally,
they provide yet another motivation to move into the quantum domain, beyond
cracking secure codes \cite{Shor94a} and efficient database queries
\cite{Grov97a}. They promise extremely high, super-classical compression of our
data. If implementations prove out, they will be valuable for improving
communication technologies.  However, they will also impact quantum computing
itself, especially for Big Data applications, as markedly less information will
have to be moved when it is quantum compressed.

\acknowledgments

We thank Ryan James, Paul Riechers, Alec Boyd, and Dowman Varn for many useful conversations. The authors thank the Santa Fe Institute for its hospitality during visits. JPC is an SFI External Faculty member. This material is based upon work supported by, or in part by, the John Templeton Foundation and the U. S. Army Research Laboratory and the U. S. Army Research Office under contract W911NF-13-1-0390.


\bibliography{chaos}
\end{document}